\documentclass[journal]{IEEEtran}
\usepackage{amsmath,graphicx}
\usepackage{graphics}
\usepackage{bm}
\usepackage{amsthm}
\usepackage{amssymb}
\usepackage{nicefrac}
\usepackage{bbold}
\usepackage{comment}
\usepackage{flushend}
\usepackage{amsfonts}
\usepackage{color}
\usepackage{cite}
\usepackage[]{algorithm2e}
\usepackage[utf8]{inputenc} 
\usepackage[T1]{fontenc}
\usepackage[nolist]{acronym}
\usepackage[export]{adjustbox}
\usepackage{comment}
\graphicspath{{figures/}}

\begin{acronym}[SPACEEEEEE]
	\acro{AWGN}{additive white Gaussian noise}
	\acro{AMP}{approximate message passing}
	\acro{BER}{bit error rate}
	\acro{BPSK}{binary phase shift keying}
	\acro{BP}{belief propagation}
	\acro{CF}{cycle-free}
	\acro{CPM}{circulant permutation matrices}
	\acro{CP}{central processor}
	\acro{CN}{check node}
	\acro{CRAN}{cloud - radio access network}
	\acro{CSA}{coded slotted ALOHA}
	\acro{CSCG}{circularly-symmetric complex Gaussian}
	\acro{CSI}{channel state information}
	\acro{CU}{central unit}
	\acro{DtF}{detect-and-forward}
	\acro{EN}{edge node}
	\acro{ESD}{empirical spectral distribution}
	\acro{FBL}{finite-blocklength}
	\acro{FEC}{forward error correction}
	\acro{F-RAN}{fog-radio access network}
	\acro{GAMP}{generalized approximate message passing}
	\acro{GQ}{generalized quadrangle}
	\acro{H-GAMP}{hybrid generalized approximate message passing}
	\acro{H-AMP}{hybrid approximate message passing}
	\acro{i.i.d.}{identical and independent distributed}
	\acro{IoT}{Internet-of-Things}
	\acro{IRSA}{irregular repetition slotted ALOHA}
	\acro{IWSN}{industrial wireless sensor network}
	\acro{JSC}{joint source-channel}
	\acro{LCR}{low-code rate}
	\acro{LDCD}{Low-density code-domain}
	\acro{LDPC}{low-density parity check}
	\acro{LDS}{low-density spreading}
	\acro{LLR}{log-likelihood ratio}
	\acro{LR}{likelihood ratio}
	\acro{LSD}{limiting spectral distribution}
	\acro{LSL}{large system limit}
	\acro{MF}{matched filter}
	\acro{MnAC}{many-access channel}
	\acro{MPA}{message passing algorithm}
	\acro{MRC}{maximum ratio combining}
	\acro{MAC}{multiple-access channel}
	\acro{ML}{maximum likelihood}
	\acro{mMTC}{massive machine-type communications}
	\acro{MUD}{multi-user detector}
	\acro{NOMA}{non-orthogonal multiple access}
	\acro{OFDM}{orthogonal frequency division multiplex}
	\acro{pdf}{probability distribution function}
	\acro{pmf}{probability mass function}
	\acro{QC}{quasi-cyclic}
	\acro{QF}{quantize-and-forward}
	\acro{QoI}{quantity of interest}
	\acrodefplural{QoI}{quantities of interest}
	\acro{RB}{resource block}
	\acro{RE}{resource element}
	\acro{RRH}{remote radio head}
	\acro{SNR}{signal-to-noise ratio}
	\acro{SCMA}{sparse coded multiple access}
	\acro{SSC}{separate source-channel}
	\acro{SUMF}{single user matched filter}
	\acro{TAKTILUS}{\textit{Taktiles Internet f\"ur sichere und zeitsensitive Anwendungen der Industrie- und Prozessautomation}}
	\acro{TBRA}{type-based random access}
	\acro{TBMA}{type-based multiple access}
	\acro{UE}{user equipment}
	\acro{VN}{variable node}
	\acro{gNB}{gNodeB}
\end{acronym}
\begin{document}
\title{Joint Source-Channel Coding for Semantics-Aware Grant-Free Radio Access in IoT Fog Networks}
\author{Johannes Dommel, 
Zoran Utkovski, Osvaldo Simeone and S\l awomir Sta\'{n}czak
\thanks{J.~Dommel (e-mail: johannes.dommel@hhi.fraunhofer.de), Z. Utkovski and S. Sta\'{n}czak are with the Department of Wireless Communications and Networks, Fraunhofer Heinrich-Hertz-Institute, Berlin, Germany. S.~Sta\'{n}czak is with the Department of Telecommunication Systems, Technical University of Berlin, Berlin, Germany. O.~Simeone is with King's Communications, Learning \& Information Processing (KCLIP) lab, CTR, Dept. of Engineering, King's College London.}
} 
\maketitle
\begin{abstract}
A \ac{F-RAN} architecture is studied for an \ac{IoT} system in which wireless sensors monitor a number of multi-valued events and transmit in the uplink using grant-free random access to multiple \acp{EN}.
Each \ac{EN} is connected to a \ac{CP} via a ﬁnite-capacity fronthaul link. 
In contrast to conventional information-agnostic protocols based on \ac{SSC} coding, where each device uses a separate codebook, this paper considers an information-centric approach based on \ac{JSC} coding via a non-orthogonal generalization of \ac{TBMA}. 
By leveraging the semantics of the observed signals, all sensors measuring the same event share the same codebook (with non-orthogonal codewords), and all such sensors making the same local estimate of the event transmit the same codeword. The \ac{F-RAN} architecture directly detects the events' values without first performing individual decoding for each device. 
Cloud and edge detection schemes based on Bayesian message passing 
are designed and trade-offs between cloud and edge processing are assessed. %
\end{abstract}
\acresetall
\IEEEpeerreviewmaketitle
\section{Introduction}
\label{sec:intro}
\begin{figure}[t!]
\centering
\includegraphics[clip, trim=0.2cm 0.2cm 0.4cm 0cm, width=1\columnwidth]{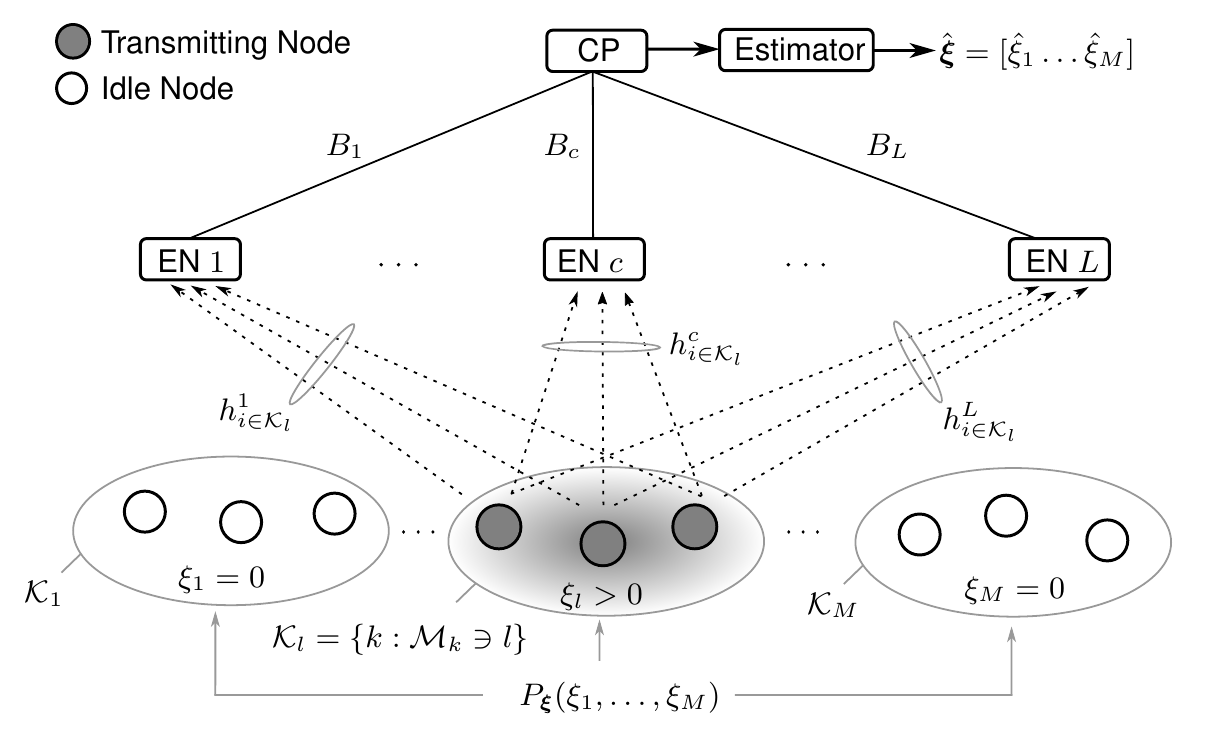}
\caption{A wireless \acf{F-RAN} for event driven random access: a set of sensors $\mathcal{K}$ monitors jointly $M$ independent events, with each event $m$ being either inactive ($\xi_m = 0$) or active with an associated scalar state value $\xi_m \in \{1,...,R\}$. Each sensor $k \in \mathcal{K}_m \subset \mathcal{K}$ measuring an active event~$m \in \mathcal{M}_k$ 
transmits a message to $L$~\acfp{EN} over a wireless fading channel. The \acp{EN} are connected via capacity-limited fronthaul links of capacity $B_1, \ldots, B_L$ to a \acf{CP} for joint decoding and estimation of the events values $\hat{\boldsymbol{\xi}} = [\hat{\xi}_1 \ldots \hat{\xi}_M]$.}
\label{fig:event_based_scenario}
\end{figure}

\IEEEPARstart{D}{ue} 
to the growing interest in \ac{IoT} applications, there has been an intense research effort on \ac{mMTC} for 5G networks and beyond~\cite{Mahmood2020, David18, chen2020massive}. 
In these networks, standard medium access control protocols that recover the individual messages of participating devices require spectral resources that scale at least linearly with the number of active users \cite{abramson70,liva11,paolini2015bcoded, Chen2017}. 
This paper proposes the integration of two distinct mechanisms that aim at reducing the communication overhead, namely (\emph{i}) the use of cloud and edge processing in \acp{F-RAN}~\cite{Park2016}; and (\emph{ii}) the application of semantics-aware medium access protocols that are designed to recover the aggregated information of interest rather than the individual messages (see, e.g.,~\cite{popovski2020semantic,kountouris2020semantics}).    
To elaborate, we consider a multi-cell \ac{F-RAN} architecture\cite{Park2016}, as illustrated in Fig.~\ref{fig:event_based_scenario}, where \ac{IoT} devices are connected to \acp{EN} in a cell-free fashion. 
Each \ac{EN} is connected via a ﬁnite-capacity fronthaul link to a \ac{CP}. 
In the system under study, multiple \ac{IoT} sensor devices measure correlated events and transmit messages in a grant-free fashion via wireless channels to the \acp{EN}.
The events may be inactive, and functions of multiple \ac{IoT} sensors' measurements, rather than individual measurements, are of interest to the receiver.
In information-theoretic terms, the problem is thus not one of channel coding in a \ac{MAC} for reliable communication of individual messages, but rather that of \ac{JSC} coding for effective inference of correlated \acp{QoI}.

\smallskip
\noindent\textit{Related work:} 
A notable instance of information-centric \ac{MAC} protocols is \ac{TBMA}\cite{mergen06}. 
With \ac{TBMA}, each measurement value for a given \ac{QoI} is assigned an orthogonal codeword, and the receiver infers the desired \ac{QoI} from a histogram at the outputs of a filter-bank matched to the codewords~\cite{mergen06, Liu04, anandkumar07}. 
A potentially more efficient solution based on a non-orthogonal generalization of \ac{TBMA} has been proposed in~\cite{dommel2019joint}. 
Accordingly, all sensors measuring the same event share the same codebook with non-orthogonal codewords, and the base station directly detects the events' values using a Bayesian message passing technique. 
Recently, \ac{TBMA} has been extended to multi-cell \acp{F-RAN} for \ac{IoT} applications under centralized or decentralized decoding in\cite{Kassab19}. 
%

\smallskip
\noindent\textit{Contribution:} 
In this paper, we study the integration of cloud detection in \ac{F-RAN} with grant-free transmission based on the semantics-aware non-orthogonal \ac{TBMA} protocol~\cite{dommel2019joint}. 
In the proposed approach, detection is performed in a centralized fashion in the cloud based on either \ac{DtF}~\cite{Utkovski17} or \ac{QF} utilizing capacity limited fronthaul links.
We design \ac{DtF} and \ac{QF} schemes based on Bayesian message passing by leveraging the \ac{H-GAMP} algorithm~\cite{rangan17}. 
Finally, we numerically evaluate the relative performance of the proposed \ac{DtF} and \ac{QF} schemes under capacity-constrained fronthaul. 
\vspace{-5pt}
\section{Event-Based Random Access for  Fog-IoT: System Model and Coding Scheme}
\label{sec:system_model}
\noindent
\textit{Scenario:} 
We consider an \ac{F-RAN} \ac{IoT} architecture consisting of a set $\mathcal{L}$ of $L$~\acp{EN}, each connected to a \ac{CP} unit via a capacity constrained fronthaul link, as illustrated in Fig.~\ref{fig:event_based_scenario}. 
In this scenario, a set $\mathcal{K}$ of $K$ devices jointly monitor a set $\mathcal{M}$ of $M$ multi-valued events. 
Each event $m$ is characterized by an independent scalar random variable ${\xi_m \in \{0, 1, \ldots, R\}}$, with $P_{\boldsymbol{\xi}}(\xi_m=0)=1-\rho$ for some $0\leq\rho\leq1$ representing the probability that event $m$ is inactive. 
When the event is active, the event variable $\xi_m$ takes one of the values in the set $\{1,\ldots,R\}$, so that parameter $R$ (or, more properly, its logarithm) measures the amount of information attached to the occurrence of an event.
Each device $k$ can simultaneously monitor a subset of events $\mathcal{M}_k\subseteq\{1,...,M\}$. Therefore, the devices can be partitioned into $M$, generally overlapping, groups ${\mathcal{K}_m = \{k\in\{1,..., K\}: \mathcal{M}_k \ni m\}}$. 

\smallskip
\noindent
\emph{Coding scheme:} Each device $k$ performs a local (real-valued) measurement $u_k$, which is, in general, correlated with all the variables $\xi_m$ for $m\in\mathcal{M}_k$. 
For each event $m\in\mathcal{M}_k$, the local measurement $u_k$ is mapped to a value ${\phi_m(u_k)\in\{0,1, \ldots, R\}}$, which is the \emph{local estimate} of event $m$. 
For transmission, each local estimate $\phi_m(u_k)$ 
is mapped by device $k$ into a codeword ${\boldsymbol{s}^m_{\phi_m(u_k)}\in \mathbb{C}^{N \times 1}}$, subject to a power constraint $ \|\boldsymbol{s}^m_{\phi_m(u_k)}\|_2^2 \leq 1$. 
The codewords for each event $m$ are selected from a shared codebook ${\boldsymbol{S}^m = [\boldsymbol{s}^m_{0}\ldots \boldsymbol{s}^m_{R}] \in \mathbb{C}^{N \times (R+1)}}$ of $R+1$ generally non-orthogonal codewords (columns). 
For future reference, we define $\boldsymbol{S} = [\boldsymbol{S}^1\ldots\boldsymbol{S}^M] \in \mathbb{C}^{N \times M(R+1)}$ to be a matrix that collects all codebooks.

\smallskip
\noindent \emph{Channel model:} We assume time synchronization and transmission over a block-fading channel model with coherence time–frequency span no smaller than that occupied by the codewords' duration. The signal received at \ac{EN} $c$ can be written as
\begin{equation}\label{eq:model1}
    \boldsymbol{y}^c=\sum_{k\in\mathcal{K}}  h^c_k \sum_{m\in\mathcal{M}_k} \boldsymbol{s}^m_{\phi_m(u_k)} + \boldsymbol{v}^c,
\end{equation}
where $h^c_k$ denote the fading coefficient for the link between device $k$ and \ac{EN}~$c$, which is assumed to be \ac{i.i.d.}~$\sim\mathcal{CN}(0, \sigma_h^2)$, and $\boldsymbol{v}^c \in \mathbb{C}^{N \times 1}$ the additive noise vector with elements \ac{i.i.d.}~$\sim\mathcal{CN}(0, \sigma_v^2)$.

\smallskip\noindent
To obtain a matrix notation, we define for each device $k$ the binary \emph{measurement} vector 
\begin{equation}
\boldsymbol{c}_k =[ (\boldsymbol{c}_k^1)^{T}\ldots(\boldsymbol{c}^M_k)^{T}]^{T} \in \{0, 1\}^{M(R+1) \times 1},
\label{eq:x}
\end{equation} 
with
\begin{align}
    \boldsymbol{c}_k^m = 
    \begin{cases}
    \boldsymbol{e}_{\phi_m(u_k)}   & \text{if}~m \in \mathcal{M}_k\\
    \boldsymbol{e}_{0} & \text{otherwise}, \label{eq:x_k_input}
    \end{cases}
\end{align}
where $\boldsymbol{e}_{r}$ is an $R+1$-dimensional binary vector with a single non-zero-entry at the $(r+1)$-th position.\footnote{$\boldsymbol{c}_k$ can be interpreted as a \emph{one-hot} encoding of all estimates at device $k$.} 
%
%

\noindent
With this definition, the received signal \eqref{eq:model1} at \ac{EN} $c$ can be described in matrix-notation as $\boldsymbol{S}\boldsymbol{x}^c + \boldsymbol{v}^c$ with
\begin{align}
\boldsymbol{x}^c &= \left(\boldsymbol{h}^c \otimes \boldsymbol{I}_{M(R+1)} \right)^T\boldsymbol{c},\label{eq:x_c}
\end{align}
where $\boldsymbol{h}^c = [h^c_1\ldots h^c_K]^T$ is the vector of channel coefficients; 
$\otimes$ the Kronnecker product; 
$\boldsymbol{I}_{M(R+1)}$ the identity matrix of size $M(R+1)$ and ${\boldsymbol{c} = [\boldsymbol{c}_1^T \ldots \boldsymbol{c}_K^T]^T}$ the stacked vector of measurements.
Note, that $\boldsymbol{x}^c$ is a (sparse) Bernoulli-Gaussian vector, where each non-zero element 
constitutes the superposition of complex-normal fading coefficients.%

\smallskip
\noindent 
\emph{Fronhthaul constraint:} We assume a packetized fronthaul transmission, e.g., via Ethernet, by considering a limited overall number of bits $B_c$ that each \ac{EN} $c$ can communicate error-free to the \ac{CP} per \emph{fronthaul use}. 

\smallskip\noindent
\emph{Error probability:} The \ac{CP} aims at estimating the state of each event ${\hat{\boldsymbol{\xi}} = [\hat{\xi}_1 \ldots\hat{\xi}_M]}$, where the average (per event) error probability is defined as
\begin{align}
    P_e &\doteq \frac{1}{M} \sum_{m \in \mathcal{M}} \text{Pr}\left\{\xi_m \neq \hat{\xi}_m\right\}\label{eq:Pe}.
\end{align}
%
%

\smallskip
\noindent
We note that the outlined \ac{MAC} protocol can be considered as a generalization of \ac{TBMA}~\cite{mergen06}, given that the latter assumes a single event, i.e. $M\doteq 1$, and the use of $R+1$ \emph{orthogonal} codewords of length $N\geq R+1$. 
%

\section{F-RAN Processing with Limited Fronthaul Capacity} 
\label{sec:edge_vd_cloud}
%
%
%
In this section, we introduce a Bayesian decoder based on \ac{GAMP}. The proposed approach extends the decoder introduced in~\cite{dommel2019joint} from single EN-detection to the \ac{F-RAN} architecture  discussed in  Section~\ref{sec:system_model}. 
We derive a graphical model and develop two fronthaul processing schemes: (\emph{i})  \ac{DtF}, whereby each \ac{EN} produces local estimates and forwards quantized soft-information to the \ac{CP}; and (\emph{ii}) \ac{QF}, whereby each \ac{EN} directly forwards a quantized version of the  received signal to the \ac{CP}. %
%

\smallskip
\noindent
\emph{Graphical Model:}
The relation between the involved \emph{random} variables, i.e. the (input) ${\mathbf{x} = [(\mathbf{x}^1)^T\ldots (\mathbf{x}^L)^T]^T}$, the (output) observations ${\mathbf{y} = [(\mathbf{y}^1)^T\ldots (\mathbf{y}^L)^T]^T}$ and the (hidden) variables $\boldsymbol{\xi}$ 
can be described at the \ac{CP} via a graphical model, where the input $\mathbf{x}$ depends on $\boldsymbol{\xi} \sim P_{\boldsymbol{\xi}}$ via the  mapping~\eqref{eq:x}-\eqref{eq:x_k_input}, which we denote as $p_{\mathbf{x}^c\vert \boldsymbol{\xi}}$.
The output $\mathbf{y}$ is generated subject to the conditional \ac{pdf} $p_{\mathbf{y}\vert \mathbf{z}}$ capturing the effect of the additive white Gaussian noise, where $\mathbf{z} = [(\mathbf{z}^1)^T \ldots (\mathbf{z}^L)^T]^T$ is the output of a (dense) linear mixing $\boldsymbol{A} \mathbf{x}$ with $\boldsymbol{A}= \boldsymbol{I}_L \otimes \boldsymbol{S}$.  
According to our transmission scheme, the $i$-th element of $\mathbf{z}^c$ is defined as ${\mathrm{z}_i^c=(\boldsymbol{a}^c_i)^T\mathbf{x}^c}$ with $\boldsymbol{a}^c_i$ being the $i$-th row of $\boldsymbol{A}^c = \boldsymbol{S}$, which is the $c$-th block of $\boldsymbol{A}$. 
In the following, we adopt the graphical model for \ac{QF} and \ac{DtF} considering a limited fronthaul capacity.
\subsection{Quantize-and-Forward}
With \ac{QF}, each \ac{EN}~$c \in \mathcal{L}$ forwards a quantized version of the received symbols $\tilde{\boldsymbol{y}}^c = \mathcal{Q}^c(\boldsymbol{y}^c)$ via the fronthaul link, and the \ac{CP} uses ${\tilde{\boldsymbol{y}} = [(\tilde{\boldsymbol{y}}^1)^T \ldots (\tilde{\boldsymbol{y}}^L)^T]^T}$ to carry out joint decoding. 
The impact of the fronthaul quantization can be modeled as Gaussian test channel \cite{Gamal12},  
such that~\eqref{eq:model1} received via the $c$-th fronthaul link from \ac{EN}~$c$ at the \ac{CP} can be written as 
${\tilde{\boldsymbol{y}}^c = \boldsymbol{y}^c + \boldsymbol{q}^c}$ 
%
where $\boldsymbol{q}^c \in \mathbb{C}^{N\times1}$ represents the quantization noise vector with elements being \ac{i.i.d.} $\mathcal{CN}(0, \sigma^2_{q^c})$~\cite{Park14}. 
Following rate-distortion arguments \cite{Cover06}, $\sigma^2_{q^c}$ is upper bounded as $\frac{P}{2^{C_c} - 1}$, with $P$ being the signal power and $C_c = B_c/N$ the fronthaul rate in \emph{bit per complex sample}.
\begin{figure}[hbt]
\centering
\includegraphics[width=1\columnwidth]{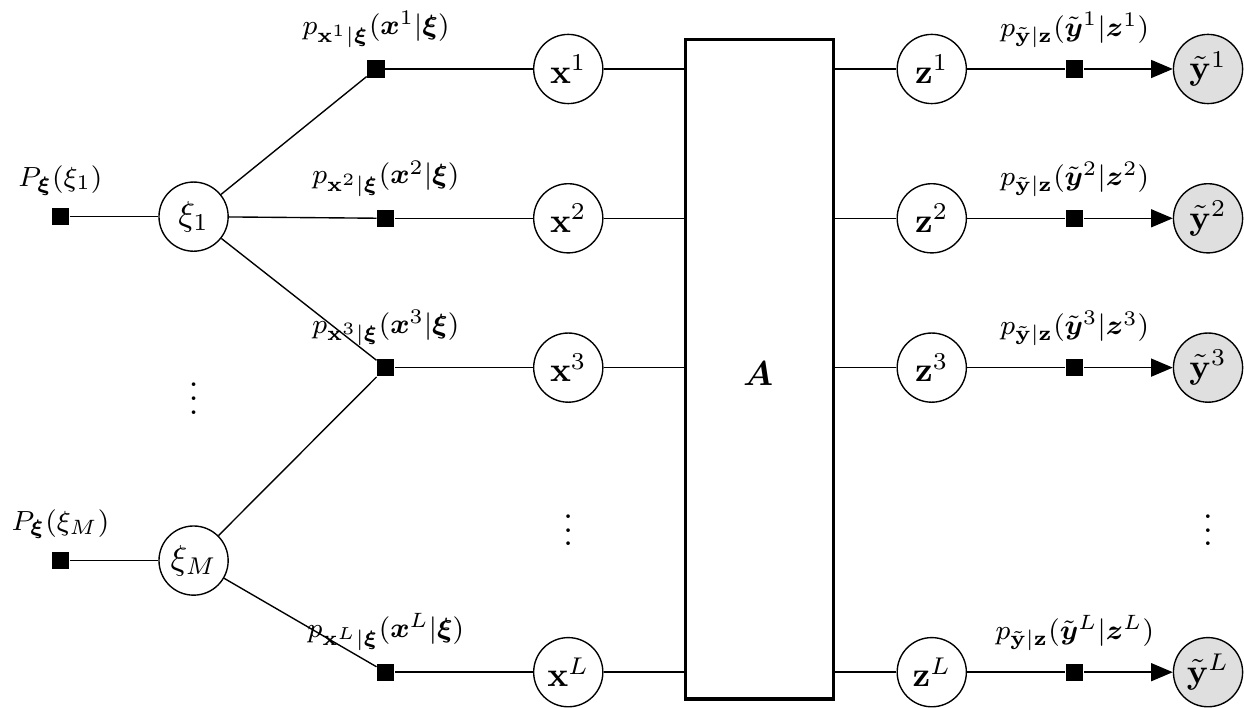}
\caption{Factor graph representation of the \acf{F-RAN} system under study with \acf{QF}. 
}  
\label{fig:graphical_model}
\end{figure}%

By the factor graph representation, see Fig.~\ref{fig:graphical_model}, the joint \ac{pdf} $p_{\boldsymbol{\xi},\mathbf{x},\tilde{\mathbf{y}}}$ of the triple $(\boldsymbol{\xi},\boldsymbol{x},\tilde{\boldsymbol{y}})$ factorizes as
\begin{align}
\prod_{m=1}^{M} P_{\boldsymbol{\xi}}(\xi_m) \prod_{j = 1}^{LM(R+1)} p_{\mathbf{x}\vert \boldsymbol{\xi}} (x_j \vert \xi_m) \prod_{i=1}^{LN} p_{\tilde{\mathbf{y}}\vert \mathbf{z}}(\tilde{y}_i\vert z_i),\label{eq:factorizationQF}
\end{align}
%
%
where the conditional \ac{pdf} $p_{\tilde{\mathbf{y}}\vert \mathbf{z}}$ captures the effect of the receiver- and quantization noise.
%

\noindent
Given the factorization \eqref{eq:factorizationQF}, the detector at the \ac{CP} aims at computing the posterior distribution $p_{\bm{\xi}\vert \tilde{\mathbf{y}}}(\bm{\xi}\vert \tilde{\boldsymbol{y}})$ of the events' state vector $\bm{\xi}$ given the quantized observation vector $\tilde{\boldsymbol{y}}$.
With \ac{QF}, the relation between the structured sparsity introduced by the hidden variables $\boldsymbol{\xi}$ on the input variables $\boldsymbol{x}$, can be exploited using the \ac{H-GAMP} algorithm~\cite{rangan17}, which provides an efficient solution with desirable empirical performance for group sparsity problems with overlapping groups. 
\ac{H-GAMP} operates by iteratively exchanging soft information between two modules: the first carries out standard \ac{GAMP} by treating the entries of the vector $\boldsymbol{x}$ as independent, while the second refines the output of the first by leveraging the correlation structure of the entries of vector $\boldsymbol{x}$. 
%
%
%
With the posterior distribution, the \ac{CP} calculates for each event the \acp{LLR}
\begin{align}
    l_{m, r} &= \ln \left( \frac{p_{\bm{\xi}|\tilde{\mathbf{y}}}(\xi_m = r|\tilde{\bm{y}})}{p_{\bm{\xi}|\tilde{\mathbf{y}}}(\xi_m = 0|\tilde{\bm{y}})}\right), 
\end{align}
associated with the belief that event $m$ is \emph{active} with value $r \in [R]$. 
%
%
The estimator then selects the decisions for $\hat{\xi}_m$ as 
\begin{align}
\hat{\xi}_m &= \begin{cases} 0 & \text{if} \quad l_{m,r} < l^{th}_{m,r},\quad \forall r \in [R]
            \\\arg\max\limits_{r\in[R]} l_{m,r} & \text{otherwise}. \end{cases}\label{eq:decision_rule_qf}
\end{align}
The thresholds $l^{th}_{m,r}$ can be selected, e.g., to minimize the Bayesian risk~\cite{kay1998fundamentals} that included individual costs for false-alarm (false positive) and missed-detection (false negative).
%
%
%
%
\subsection{Detect-and-Forward}
\label{sec:DtF}
With \ac{DtF}, each \ac{EN} $c\in \mathcal{L}$ performs \emph{local detection} and forwards quantized soft-information to the \ac{CP}. 
The \ac{CP} then fuses the local decisions to obtain the final estimates for each event.
For detection at each \ac{EN}~$c$, the joint \ac{pdf} $p_{\boldsymbol{\xi},\mathbf{x}^c,\mathbf{y}^c}(\cdot,\cdot,\cdot)$ of the triple 
$(\boldsymbol{\xi},\boldsymbol{x}^c,\boldsymbol{y}^c)$ factorizes as
\begin{align}
\prod_{m=1}^{M} P_{\boldsymbol{\xi}}(\xi_m) \prod_{j=1}^{M(R+1)} p_{\mathbf{x}^c\vert \bm{\xi}} (x^c_j \vert \xi_m) \prod_{i=1}^{N} p_{\mathbf{y}^c\vert \mathbf{z}}(y^c_i\vert z_i^c).
\label{eq:factorization}
\end{align}

\smallskip \noindent 
Using \eqref{eq:factorization}, the local detector at \ac{EN} $c$ aims at computing the posterior distribution $p_{\bm{\xi}\vert \mathbf{y}^c}(\bm{\xi}\vert \boldsymbol{y}^c)$ of the events' state vector $\bm{\xi}$ given the local observation $\boldsymbol{y}^c$. 
This can be done by applying the standard \ac{GAMP} operating on $\boldsymbol{A}^c = \boldsymbol{S}$. 

\smallskip
\noindent
Given the posterior distribution, each \ac{EN} $c \in \mathcal{L}$ computes the local \acp{LLR} for all $r \in [R]$ and $m \in \mathcal{M}$ as 
\begin{align}
    l^c_{m, r} &= \ln \left( \frac{p_{\bm{\xi}|\mathbf{y}^c}(\xi_m = r|\bm{y}^c)}{ p_{\bm{\xi}|\mathbf{y}^c}(\xi_m = 0|\bm{y}^c)} \right),
    \label{eq:log_dtf}
\end{align}
associated with the (local) belief at \ac{EN}~$c$, that the event variable $\xi_m$ is active with value $r$. 
For transmission over the capacity-constraint fronthaul, each \ac{EN}~$c$ applies a quantization function $\tilde{x} = \mathcal{U}^c(x)$, to the \acp{LLR}~\eqref{eq:log_dtf} according to the fronthaul bit-budget of $B_c / M$ bit \emph{per event} for all $m \in \mathcal{M}$. 
At the \ac{CP}, the beliefs are reconstructed and merged to obtain
\begin{equation}
\tilde{l}_{m,r} =\sum_{c\in\mathcal{L}}\tilde{l}_{m,r}^c,\:\: r \in [R], \label{eq.dtf_llr}
\end{equation}
which is associated with the (global) belief that the event variable $\xi_m$  is active with value $r \in [R]$.
Finally, the \ac{CP} estimates the event activity variable by comparing \eqref{eq.dtf_llr} against thresholds, c.f.~\eqref{eq:decision_rule_qf}.
%
We note that in the case of \ac{DtF}, an optimal compression performance can be achieved by using an entropy quantizer operating on the $MR$-dimensional vector of \acp{LLR}.
%
\section{Numerical Results}
\label{sec:numerical_results}
We assume a dense network with $K=80$ devices observing in total $M=8$ events with $R = 4$ values in an \ac{F-RAN} deployment with $L = 4$ \acp{EN} and fronthaul bit budget $B_c = B$, for all fronthaul links between the \acp{EN} $c \in \mathcal{L}$ and the \ac{CP}.
Each event has an activation probability $\rho = 0.1$ and the devices are configured such that each individual device observes only one of the events and that the total number of devices is partitioned into $M$ non-overlapping sets $\{\mathcal{K}_m\}$, each of cardinality $10$. 
The variance of the channel coefficients, which we recall are unknown to the transmitter devices and the receiver, is set to $\sigma^2_h = 1$. 
To increase the energy efficiency, each device $k$ is configured for transmission, \emph{only} if the locally observed event $m \in \mathcal{M}_k$ is \emph{active}, i.e., if $\phi_m(u_k) > 0$. 
The signatures of the shared codebook $\boldsymbol{S}$ of length $N$ are generated randomly with entries being \ac{i.i.d.} $\sim \mathcal{CN}(0, 1/N)$. 
We note that the convergence of \ac{AMP} for this codebook has been studied rigorously in the asymptotic limit~\cite{rangan17}. %
The average \ac{SNR} is defined \emph{per user} as $\textrm{SNR} \doteq 1/\sigma_v^2$. %
For \ac{DtF}, only the \acp{LLR} associated with the non-zero entries of the estimated event activity pattern are quantized and forwarded to the \ac{CP}. 
The threshold for \ac{DtF} and \ac{QF} is chosen to minimize the error probability $P_e$, according to~\eqref{eq:Pe}. 
%
%

\begin{figure}[htb]
\centering
\includegraphics[width=0.9\columnwidth]{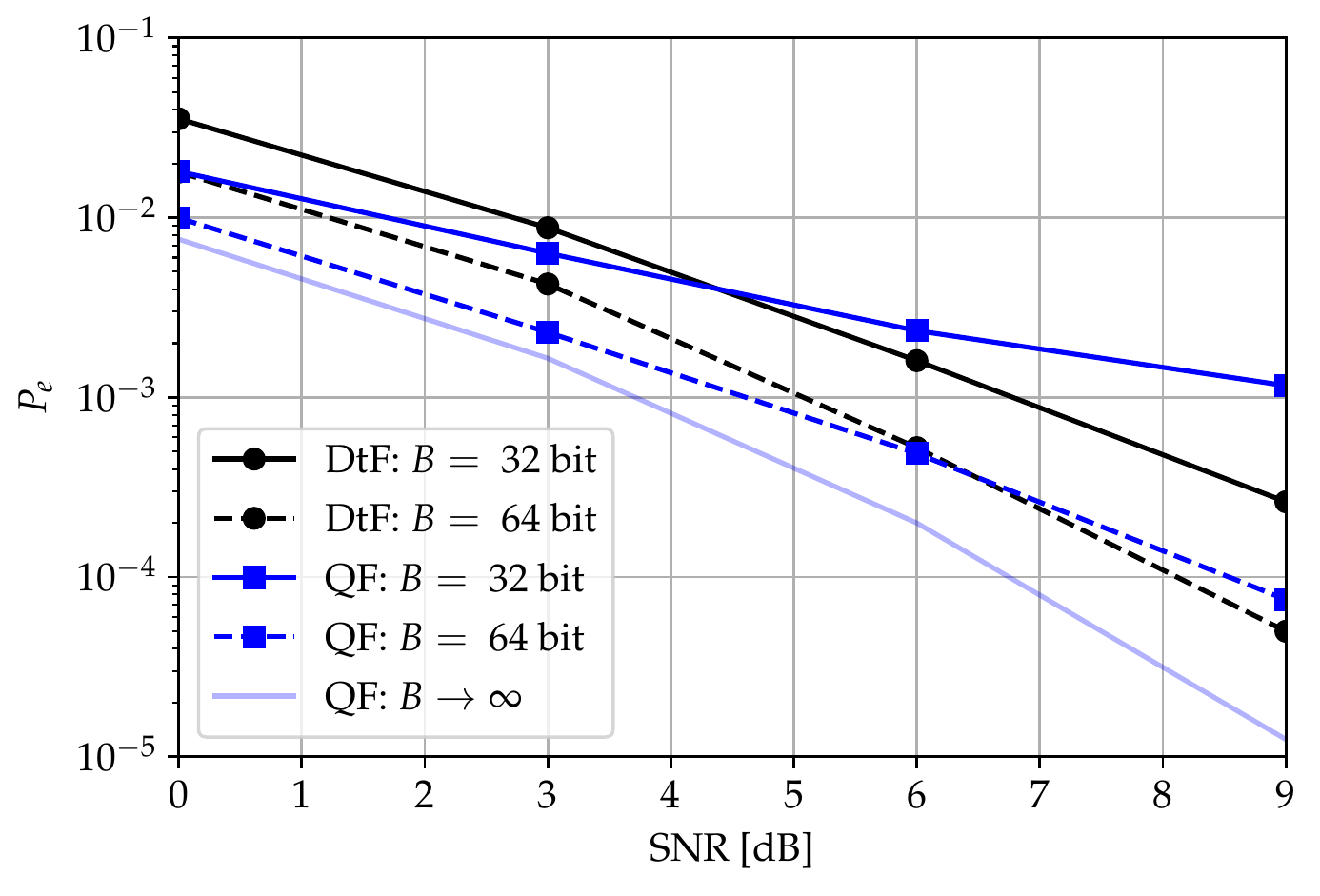}
\caption{Error rate versus \ac{SNR} for \ac{F-RAN} deployment with $L = 4$ \acp{EN}, signature length $N = 16$, and limited fronthaul capacity $B$.}  
\label{fig:error_vs_snr}
\end{figure}

\noindent
The impact of fronthaul quantization on the error rate $P_e$ is plotted in Fig.~\ref{fig:error_vs_snr} as a function of the \ac{SNR} for both fronthaul processing schemes under different fronthaul bit budgets.
\ac{DtF} is seen to outperform \ac{QF} in the regime of high \ac{SNR}, with crossing point occurring at lower \ac{SNR} levels for a small fronthaul budget.
This is because, with a sufficiently large \ac{SNR} and small enough fronthaul capacity, the potential advantages of centralized detection at the \ac{CP} are offset by the fronthaul quantization noise, and local detection is preferable.

\begin{figure}[htb]
\centering
\includegraphics[width=0.9\columnwidth]{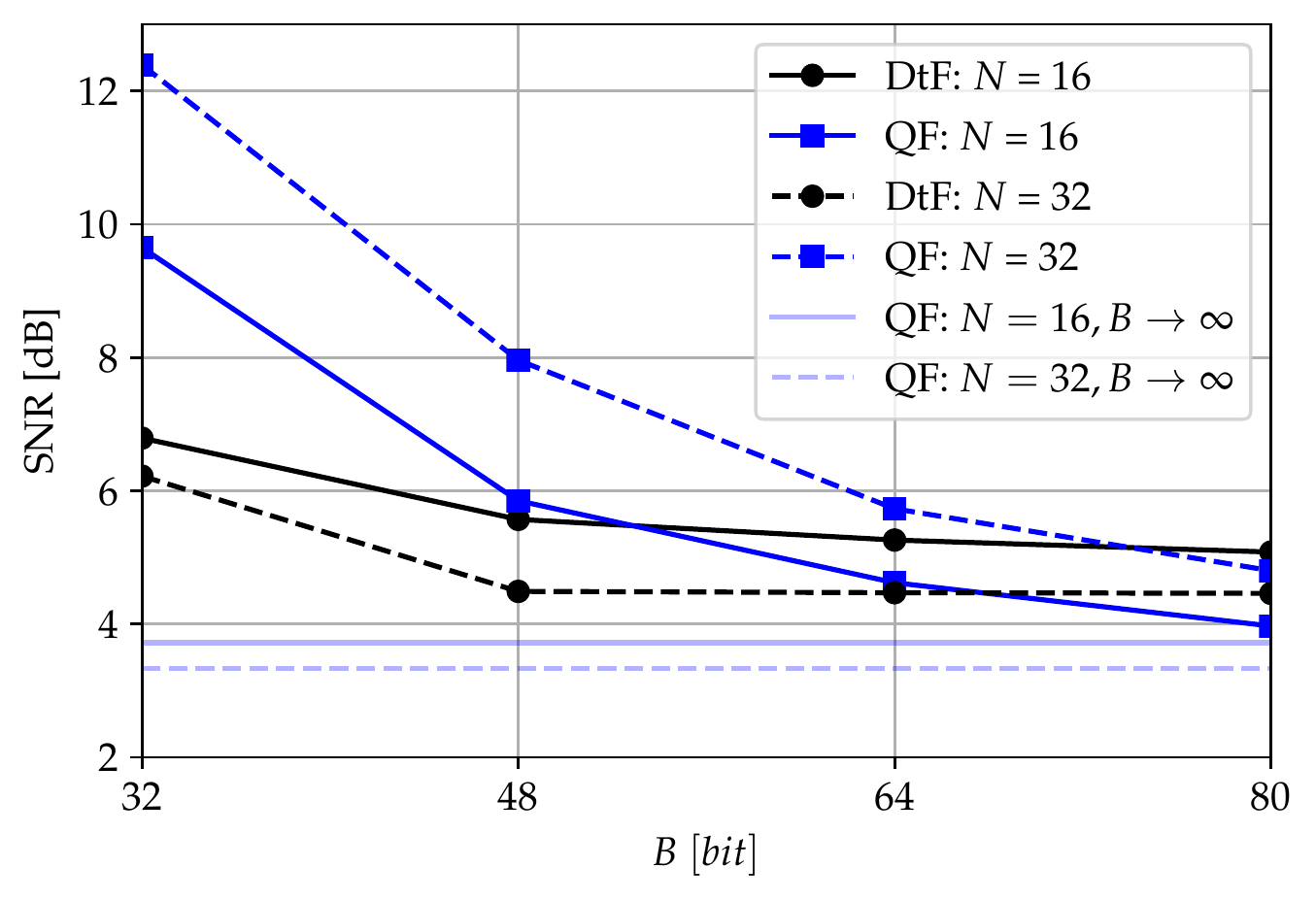}
\caption{Required \ac{SNR} to achieve a target reliability $P_e \leq 10^{-3}$ as a function of the fronthaul link budget $B$ and signature length $N$.}  
\label{fig:cap_vs_nEn}
\end{figure}

\noindent
The comparison depends also on the length $N$ of the signature. 
To elaborate on this point, Fig.~\ref{fig:cap_vs_nEn} plots the \ac{SNR} required to meet a predefined reliability target $P_e \leq 10^{-3}$ as a function of the fronthaul bit budget $B$ and the signature length $N$. 
As discussed, \ac{QF} is preferable only at sufficiently large fronthaul capacity levels, and the required fronthaul capacity increases with the signature length $N$.
In fact, for \ac{QF}, in the presence of stringent fronthaul constraints, it is beneficial to trade signature length for quantization  precision.

\begin{figure}[htb]
\centering
\includegraphics[width=0.9\columnwidth]{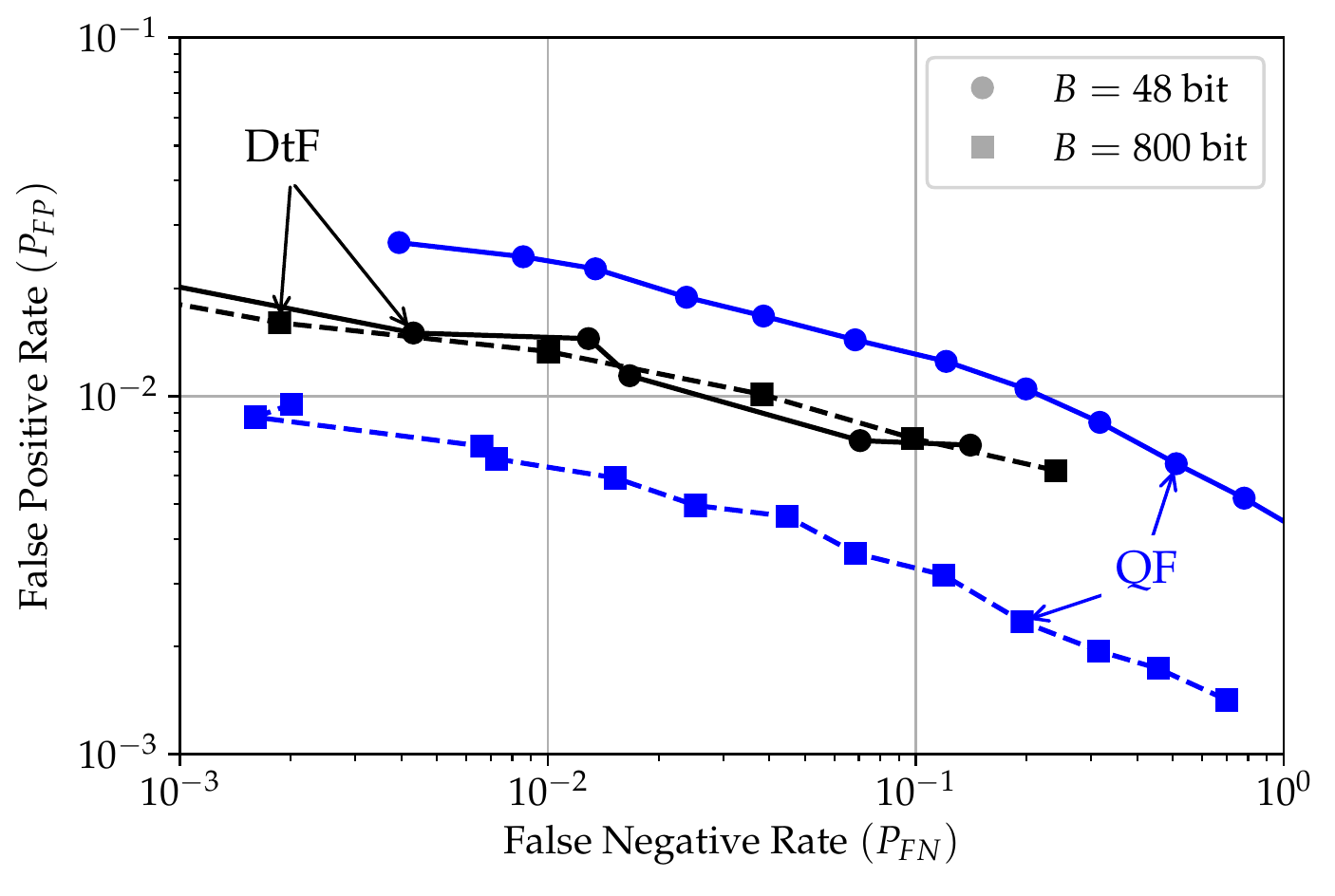}
\caption{Trade-off between false positive and false negative rates for different decision thresholds for $N = 32$ and $\text{SNR} = 0~\text{dB}$.} \label{fig:FA_vs_Pe}
\end{figure}

\smallskip
\noindent Finally, in Fig.~\ref{fig:FA_vs_Pe} we analyze the trade-off between false positive 
rate, $P_{FP} \doteq \frac{1}{M} \sum_{m = 1}^{M} \text{Pr}\{\hat{\xi}_m \neq 0 \vert \xi_m = 0 \}$, and false negative 
rate, $P_{FN} \doteq \frac{1}{M} \sum_{m = 1}^{M} \text{Pr}\{\hat{\xi}_m = 0 \vert \xi_m \neq 0 \}$, 
obtained by varying the decision threshold.
In line with the discussion so far, \ac{QF} is seen to provide significant advantages when the fronthaul capacity $B$ is sufficiently large. 
In contrast, the fronthaul requirement of \ac{DtF} are more modest, but the performance of \ac{DtF} is constrained by the limitations of local detection.
\section{Conclusions}
\label{sec:summary_discussion}
This paper has introduced a semantics-aware protocol for event-driven grant-free access in \ac{IoT} \acp{F-RAN} with fronthaul capacity constraints. 
The protocol adopts a joint source-channel coding scheme, based on a non-orthogonal generalization of \ac{TBMA}, that directly detects the quantities of interest and  yields spectral requirements that scale with the number of events to be monitored rather than with the number of devices. The power and spectral requirements are further improved through integration with cloud- and edge detection based on Bayesian message passing. 
We have evaluated numerically the relative performances of edge-cloud processing based on \ac{DtF} and \ac{QF}, and assessed the trade-offs between the codeword length, the fronthaul capacity and the required \ac{SNR}, given a  predefined reliability target. 
A general observation is that DtF outperforms QF in the presence of stringent fronthaul constraints, with  the effect being more pronounced for higher SNR values. In this operational regime, the potential advantages of centralized detection at the \ac{CP} are offset by the fronthaul quantization noise, and local detection is preferable.
Further, in line with \cite{dommel2019joint} we conclude that the proposed scheme offers a significantly higher spectral efficiency as compared to conventional \ac{TBMA} by exploiting (\emph{i}) sparse activation and (\emph{ii}) structural dependencies between the variables for $L>1$.
\bibliographystyle{IEEEbib}
\bibliography{IEEEabrv, refs}
\end{document}